\documentclass[aps,pra,reprint,superscriptaddress]{revtex4-2}
\usepackage{graphicx}
\usepackage{dcolumn}
\usepackage{hyperref}
\usepackage{amsmath}
\usepackage{amssymb}
\usepackage{braket}
\usepackage{mathtools}
\usepackage{upgreek}
\usepackage[version=4]{mhchem}
\usepackage[binary-units]{siunitx}[=v2]
\usepackage{booktabs}
\DeclareSIUnit{\torr}{Torr}
\DeclareSIUnit{\bar}{bar}

\usepackage[utf8]{inputenc}
\usepackage[T1]{fontenc}

\begin{document}

\title{Electromagnetically induced transparency and optical pumping in the hyperfine Paschen-Back regime}

\author{Roberto Mottola}
\email{roberto.mottola@unibas.ch}
\author{Gianni Buser}
\author{Philipp Treutlein}
\email{philipp.treutlein@unibas.ch}
\affiliation{Departement Physik, Universit\"{a}t Basel, Klingelbergstr. 82, 4056 Basel, Switzerland.}

\date{December 19, 2023} 

\begin{abstract}
	We report spectroscopy experiments of rubidium vapor in a high magnetic field under conditions of electromagnetically induced transparency (EIT) and optical pumping. The \SI{1.1}{\tesla} static magnetic field decouples nuclear and electronic spins and shifts each magnetic state via the Zeeman effect, allowing us to resolve individual optical transitions of the D$_2$ line in a Doppler-broadened medium. By varying the control laser power driving one leg of a spectrally isolated $\Lambda$ system we tune the vapor from the EIT regime to conditions of Autler-Townes line splitting. The resulting spectra conform to simple three-level models demonstrating the effective simplification of the energetic structure. Further, we quantify the viability of state preparation via optical pumping on nuclear spin-forbidden transitions. We conclude that the ``cleanliness'' of this system greatly enhances the capabilities of quantum control in hot vapor, offering advantages in a broad variety of quantum applications plagued by spurious light-matter interaction processes, such as atomic quantum memories for light.
\end{abstract}

\maketitle

\section{\label{sec:intro}Introduction}

Since the advent of laser cooling, experiments in atomic, molecular, and optical physics have benefited from an unprecedented degree of control over matter \cite{Chu2002}. With natural linewidth limited atomic transitions and comparably narrow lasers, it is easily possible to lift energetic degeneracies in small fields and exert precise control over quantum states when atoms are cold. Generally speaking, the property that enables this is resolution of individual transitions, i.e., the energetic splitting between next nearest transitions is significantly greater than the transitions' linewidths. In hot vapor, on the other hand, atomic lines are inhomogenously broadened by atomic motion, and commonly further subject to collisional broadening \cite{Allard1982}. At room temperature, spectra in vapor cells typically have convolutional linewidths of at least \SI{500}{\mega\hertz} \cite{Pizzey2022}. This prevents cold-atom-level control over matter, as even the hyperfine structure of atomic excited states is often lost to line broadening \cite{Auzinsh2009}. On some alkali-metal lines, polarization selection rules can yield desirable restrictions on possible light-matter interactions in hot vapor \cite{Zhang2014}, but at the same time restrict the feasible operating regimes of applications such as atomic single-photon sources \cite{Walther2007} or quantum memories \cite{Buser2022}. By applying sufficiently high magnetic fields, however, another approach presents itself. The Zeeman effect lifts energetic degeneracy, and through further decoupling of the atom's nuclear spin states individual transitions once again become well resolved. This hyperfine Paschen-Back (HPB) regime is, therefore, a promising arena for atomic physics in hot vapor, as a significant hurdle to direct optical manipulation of atomic quantum states is removed.

Electromagnetically induced transparency (EIT) \cite{Fleischhauer2005} and the related but distinct \cite{Anisimov2011,Wu2022} phenomenon of Autler-Townes splitting (ATS) \cite{CohenTannoudji1996} have been extensively studied in low and zero fields. These effects have wide ranging practical uses, for instance, in precision metrology \cite{Holloway2017}, quantum memories \cite{Rastogi2019}, and tailored photon generation \cite{Du2008}. Moreover, multi-level atomic structure can produce efficient non-linear effects \cite{Zhang2007}.
In high magnetic fields, EIT has been investigated in ladder-schemes \cite{Whiting2016}, including Rydberg based ones \cite{Ma2017,Naber2017}, and V-scheme configurations \cite{Higgins2021}, as well as in a diamond-scheme in the context of four-wave mixing \cite{Whiting2018}.
Furthermore, studies of EIT in a $\Lambda$-scheme at intermediate fields (up to \SI{170}{\milli\tesla} for $^{85}$Rb) have been reported on in Ref.~\cite{Sargsyan2012}.
Moreover, a detailed investigation of optical pumping in Cs vapor at tesla-order magnetic fields, including the effect of forbidden transitions, was performed by Olsen \emph{et al.} in Ref.~\cite{Olsen2011}. Nevertheless, a thorough study of $\Lambda$-EIT/ATS in a hot, high optical depth ensemble deeply in the HPB regime is a critical prerequisite for putting this system to use in applications such as quantum memories, and has so far been outstanding.

In this article we study hot $^{87}$Rb vapor in a tesla-order magnetic field, with the purpose to isolate a three-level system in the atomic energy structure of either D line. We investigate EIT and atomic polarizability on the D$_2$ line at \SI{780}{\nano\meter} in this regime, characterizing the suitability of this system for quantum technological applications beyond sensing. In a concurrent article \cite{Mottola2023PRL}, we put our conclusions to the test with a quantum memory experiment in a microfabricated atomic vapor cell.

\section{\label{sec:theory}Hyperfine Paschen-Back Regime}
We consider the D lines of alkali-metal atoms, which are frequently used to study light-matter interactions. 
Both the ground as well as the excited states present a hyperfine structure, which is usually only partially resolved in the simple absorption spectra, and the multiple magnetic sublevels are generally unresolved.
Here we mainly focus on the $^{87}$Rb D$_2$ line.

The Hamiltonian for an alkali-metal atom in an external static magnetic field is given by
\begin{equation}
	\label{eq:Hamiltonian}
	\hat{H} = \hat{H}_0 + \hat{H}_{\text{hfs}} + \hat{H}_Z. 
\end{equation}
The first term $\hat{H}_0$ describes the coarse atomic structure, and $\hat{H}_{\text{hfs}} = A_{\text{hfs}} \, \mathbf{\hat{I}} \cdot \mathbf{\hat{J}} + \hat{H}_{\text{qp}}$ describes the hyperfine coupling, with the magnetic dipole constant $A_{\text{hfs}}$ and the electric quadrupole Hamiltonian $\hat{H}_{\text{qp}}$, which is non-zero for the $5\,^2\text{P}_{3/2}$ term. The last term of Eq.~\eqref{eq:Hamiltonian},
\begin{equation*}
	\hat{H}_Z = \mu_B \left( g_J\hat{\mathbf{J}} + g_I\hat{\mathbf{I}} \right) \cdot \mathbf{B}
\end{equation*}
describes the Zeeman interaction of the atom with a magnetic field. 
In this equation $g_J$ and $g_I$ are the $g$-factors for the total angular momentum of the electron and the nucleus, respectively. 

By applying an external magnetic field the degeneracy of the hyperfine states can be lifted through the energy shift induced by the Zeeman effect. This splitting becomes larger as a function of the strength of the magnetic field $\mathbf{B}$. 
For high magnetic fields, where the energy shift induced by the Zeeman interaction $\Delta E_{\text{Z}}$ becomes larger than the one caused by the hyperfine interaction $\Delta E_{\text{hfs}}$, we enter the HPB regime.
Generally speaking, the condition 
\begin{equation}
	\label{eq:HPBregime}
	B \gg B_0 = A_{\text{hfs}}^{\text{GS}}/\mu_B
\end{equation}
delineates between regimes \cite{Corney2006}. Here the ground-state hyperfine magnetic dipole constant $A_{\text{hfs}}^{\text{GS}}$ is used in the condition as the ground states experience a larger hyperfine splitting compared to the excited states, ensuring that all atomic levels are in the HPB regime.

For atomic states with a low principal quantum number $n$, the Hamiltonian describing the interaction of the atoms with an external magnetic field can thus be reduced to $\hat{H}_Z$ if condition \eqref{eq:HPBregime} is fulfilled.
In this limit the splitting between hyperfine levels grows linearly with $\Delta E_Z = \left( g_J m_J + g_I m_I \right)\mu_B B$, and a change of \SI{1}{\tesla} induces a frequency shift of $\pm\SI{14}{\giga\hertz}$ of the ground state sublevels. 

\begin{figure}
	\includegraphics[width=\columnwidth]{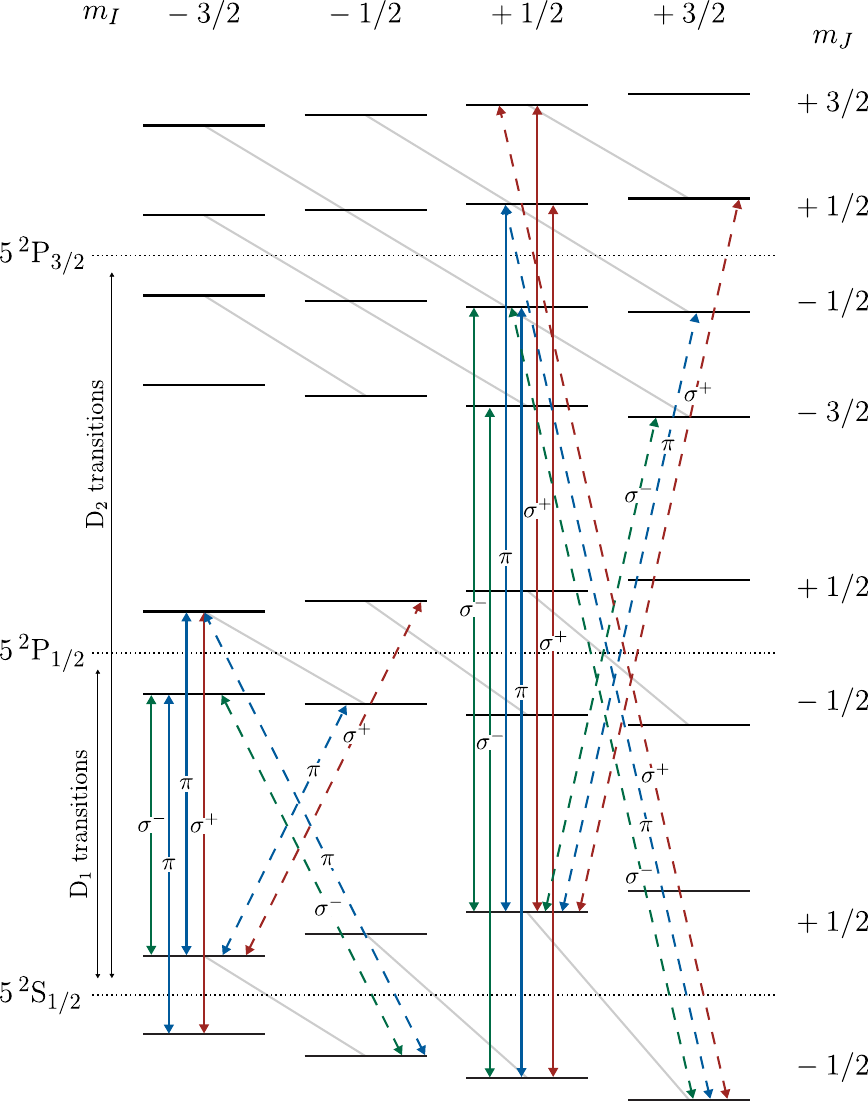}
	\caption{\label{EnergyLvls} Energy levels of $^{87}$Rb in the HPB regime represented in the $\ket{m_J,m_I}$ basis. The light gray lines indicate the coupling between the sublevels due to the hyperfine coupling. The allowed transitions for one $m_I$ manifold per D line are shown as solid lines. The ``singly forbidden'' transitions arising from the residual coupling of the ground states are represented as dashed lines. The transitions coupling to the ground states involved in the superpositions with $m_F=-1$ and $m_F=1$ in the weak field scenario are visualized for the D$_1$ and the D$_2$ line, respectively. The energy splittings are not to scale.} 
\end{figure}

\begin{figure*}
	\begin{center}
		\includegraphics[scale=1]{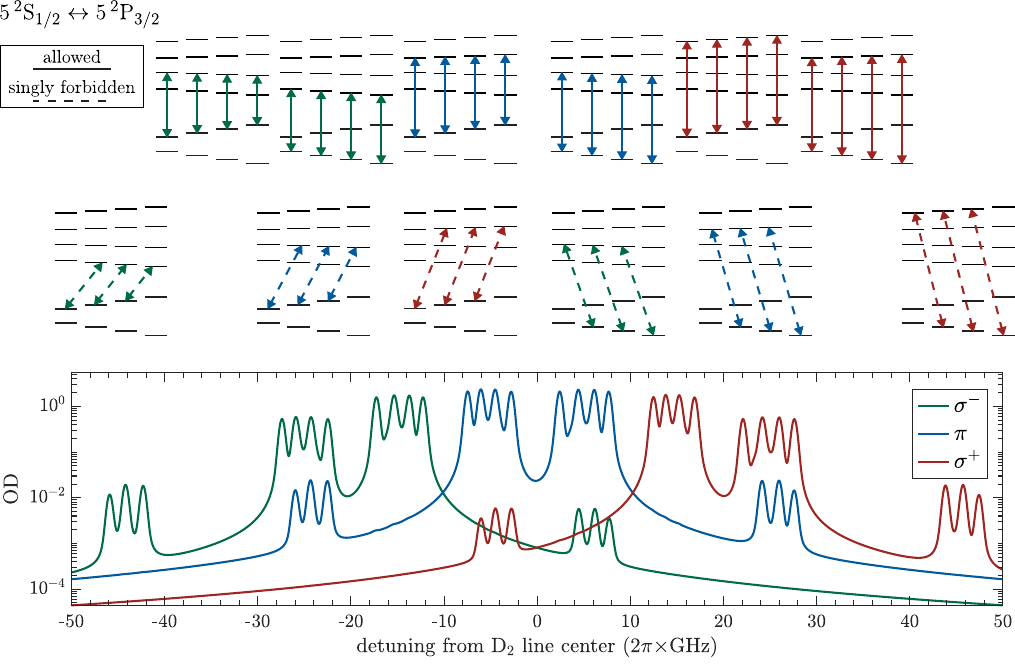}
	\end{center}
	\caption{\label{TransitionsD2} Computed spectrum of the $^{87}$Rb D$_2$ line in a \SI{1.06}{\tesla} external magnetic field. The spectrum is simulated for a \SI{2}{\milli\meter} thick cell, with $\SI{90}{\percent}$ enriched $^{87}$Rb, \SI{11}{\milli\bar} of Ar buffer gas, and an atomic temperature of \SI{97}{\degreeCelsius}. The dipole-allowed transitions appear as quadruplets, while the singly forbidden transitions form manifolds of three. The insets in the upper half of the figure show the sublevels involved in the corresponding transition manifold. The insets are roughly arranged according to the transition frequency and optical depth (OD). Solid and dashed lines illustrate allowed and singly forbidden transitions, respectively. The polarization of the transitions is color coded according to the legend.} 
\end{figure*}

In the HPB regime the nuclear spin $\mathbf{I}$ and the total angular momentum of the electron $\mathbf{J}$ decouple.
Consequently, the eigenvalue $F$ of the total angular momentum $\mathbf{F = J + I}$ of the atom and its projection $m_F$ do not represent a good choice of quantum numbers to describe the system anymore. A convenient representation is in the $\ket{m_J,m_I}$ basis.
For every value of $m_J$ there are $2I + 1$ $m_I$-sublevels (see Fig.~\ref{EnergyLvls}).
Every state of the coupled basis can be represented as a linear combination of states in the uncoupled basis according to
\begin{equation*}
\ket{F,m_F} = \sum_{\mathclap{\substack{m_J,\\ m_I = m_F - m_J}}} C^{m_F}_{m_J,m_I}\ket{m_J,m_I}.
\end{equation*}
At zero field, the constants $C^{m_F}_{m_J,m_I}$ correspond to the Clebsh-Gordan coefficients. With increasing magnetic field one of the coefficients in the sum tends to unity while the others tend to zero. The states involved in each superposition are connected by gray lines in Fig.~\ref{EnergyLvls}.  

In Fig.~\ref{EnergyLvls} the energy levels of the $^{87}$Rb $5\,^2\text{S}_{1/2}$, $5\,^2\text{P}_{1/2}$, and $5\,^2\text{P}_{3/2}$ states in the HPB regime are depicted with selected transitions. In this representation the allowed optical transitions are all vertical with $\Delta m_J = 0, \pm 1$ corresponding to $\pi$ or $\sigma^{\pm}$ polarized light, respectively.
To a first approximation, the decoupling of $\mathbf{J}$ and $\mathbf{I}$ implies that optical transitions are only allowed between states within the same $m_I$-manifold, as the light does not couple to the nuclear spin. Together with the induced energy splittings, which are much larger than the Doppler-broadened linewidth of the vapor, a clean three-level system can be addressed.

Note that, even restricting the discussion to ground states, the required order of magnitude of magnetic field strength for the Zeeman interaction to dominate varies widely.
As an extreme example, consider that for the $^7$Li D lines even the fine Paschen-Back regime, where the Zeeman interaction triumphs over the atomic fine structure, can be investigated with fields $<\SI{1}{\tesla}$ \cite{Umfer1992}. In contrast, to realize such conditions in Rb, a magnetic field of \SI{218}{\tesla} would be required \cite{Weller2012}.

For any finite magnetic field, the states actually remain a superposition.
Even at an applied field of \SI{1}{\tesla}, some residual coupling between $\mathbf{J}$ and $\mathbf{I}$ persists in the $5\,^2\text{S}_{1/2}$ ground state (see, for instance, Ref.~\cite{Mottola2023} for a precise quantification of the basis state superpositions). 
Indirect interaction of light with the nucleus through this residual $\mathbf{J}$-$\mathbf{I}$ coupling allows transitions with $\Delta m_I \neq 0$ to take place. 
We refer to transitions with $\lvert \Delta m_I \rvert = 1$ as ``singly forbidden''. They appear in manifolds of three in the spectrum and have a transition strength about 50-times weaker than the allowed transitions ($\pi$-polarization)  at \SI{1}{\tesla} field strength. 
Following conservation of total angular momentum these transitions obey the relations $\Delta m_I + \Delta m_J = 0, \pm 1$ for $\pi$ and $\sigma^{\pm}$ polarized light, respectively. A complete representation of all allowed and singly forbidden transitions for the $^{87}$Rb D$_2$ line can be found in the upper half of Fig.~\ref{TransitionsD2}.

Figure~\ref{TransitionsD2} shows the computed spectrum of the D$_2$ transitions for a vapor cell with properties similar to those of the cell we used for the measurements presented below. The spectrum is computed analogously to how it is described in Refs.~\cite{Zentile2015,Keaveney2018}.
The energy-level insets above the spectrum show the states involved in the corresponding line manifold, and are roughly ordered by transition frequency and strength. 

\section{\label{sec:setup}Experimental Apparatus}
A Bruker B-E 10 electromagnet is used to generate the static, tesla-order magnetic field perpendicularly to the propagation axis of the light. In this geometry, choosing the direction of the magnetic field as our quantization axis, linearly horizontally polarized light in the laboratory frame corresponds to $\pi$ polarization. On the other hand, pure $\sigma^+$ or $\sigma^-$ polarized light, corresponding to right- and left-circularly polarized light in the plane perpendicular to the magnetic field, respectively, are not valid polarizations for a transverse wave traveling along the optical axis.
Vertically polarized light in the laboratory frame can, however, be decomposed into an equal superposition of $\sigma^+$ and $\sigma^-$ polarizations. Experimentally, this means that these two polarizations can only be applied simultaneously. Due to the large frequency separation of $\sigma^+$ and $\sigma^-$ transitions (see Fig.~\ref{TransitionsD2}), this does not pose a problem in addressing single transitions.

\begin{figure}
	\includegraphics[scale=1]{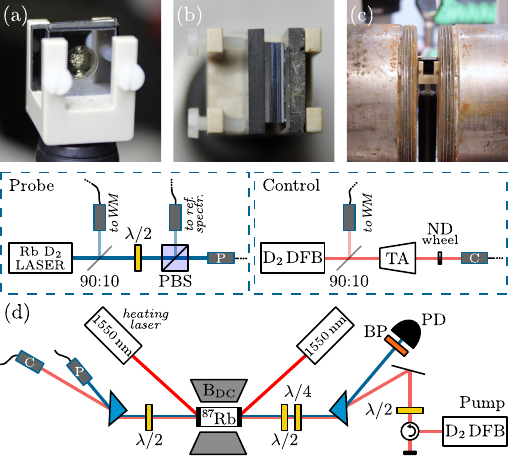}
	\caption{\label{Setup} (a) Front and (b) top view of the vapor cell used in the experiments. The cell is sandwiched by two \SI{2}{\milli\meter}-thick pieces of RG9 filter glass. (c) The ferromagnetic cores of the electromagnet constrain the optical and spatial access around the cell. (d) Schematic of the experimental spectroscopic setup. DFB, distributed feedback (laser); 90:10, beam sampler with the specified splitting ratio; WM, wavelength-meter; PBS, polarizing beam-splitter; ND, neutral density (filter); TA, tapered amplifier; B$_{\text{DC}}$, electromagnet; $\lambda/2$, half-wave plate; $\lambda/4$, quarter-wave plate; BP, bandpass (filter); PD, photo diode. The labels $P$ and $C$ indicate the optical fiber connections for probe and control beams, respectively.}
\end{figure}

A microfabricated vapor cell with a $\SI{2}{\milli\meter}$ internal thickness and a \SI{5}{\milli\meter} diameter aperture is used for the experiments [see Fig.~\ref{Setup}(a)]. 
The fabrication of this type of cell is described in Refs.~\cite{DiFrancesco2010,Pellaton2012}.
The cell is filled with $\leq \SI{90}{\percent}$ isotopically enriched $^{87}$Rb and about \SI{11}{\milli\bar} of Ar buffer gas.
To heat the cell we rely on infrared lasers and absorptive colored filter glass (Schott RG9), which is highly transmissive at the Rb D-line wavelengths.
The vapor cell is sandwiched between two $\SI{2}{\milli\meter}$ thick pieces of this filter glass, as shown in Fig.~\ref{Setup}(b). Each side is illuminated by a low-cost, multimode, telecom laser (Seminex 4PN-108).
With this technique the atoms can be heated to temperatures $> \SI{130}{\degreeCelsius}$. The atomic temperature is determined spectroscopically. 
This heating technique has previously been successfully implemented in vapor based magnetometers, where its efficiency was optimized to allow for low-power operation \cite{Mhaskar2012}. In our setup no effort was made so far to render the heating more efficient, and on the order of \SI{1}{\watt} of optical power is required to reach operational temperatures.

The spectroscopic setup used for the measurements presented below is shown in Fig.~\ref{Setup}(d).
Widely tunable external cavity diode and distributed feedback (DFB) lasers on either D line are used as weak probes.
The control light is generated by a DFB laser that is amplified with a tapered amplifier (TA). The resulting optical power gives us the option to explore a large range of Rabi frequencies. A neutral density gradient wheel is used to vary the control power. The control beam is focused to a $1/e^2$-diameter of approximately \SI{550}{\micro\meter} in the center of the cell.

Probe and control beams are combined on a polarizing calcite prism before the vapor cell. The overlap of the beams is ensured by coupling each of them into the same (auxiliary) single-mode fiber and regularly optimizing the alignment.
A second calcite prism is used to discriminate the strong control beam from the probe by polarization. At least 8 orders of magnitude of suppression can be achieved this way.
Finally, the probe is detected by a photo diode, equipped with a bandpass filter to remove ambient light.

For state preparation a further DFB laser is dedicated to optical pumping. It is aligned to be counter-propagating with respect to the control beam and is coupled in through a Faraday circulator.

The frequencies of the various lasers are referenced with a wavelength meter (HighFinesse WS-7) through auxiliary fiber ports. Furthermore, a portion of the probe light is branched off directly after the laser for a Doppler-free saturation spectroscopy to yield an absolute frequency reference. 

\section{\label{sec:Spectroscopy}Absorption Spectroscopy}
\begin{figure}
	\includegraphics[scale=1]{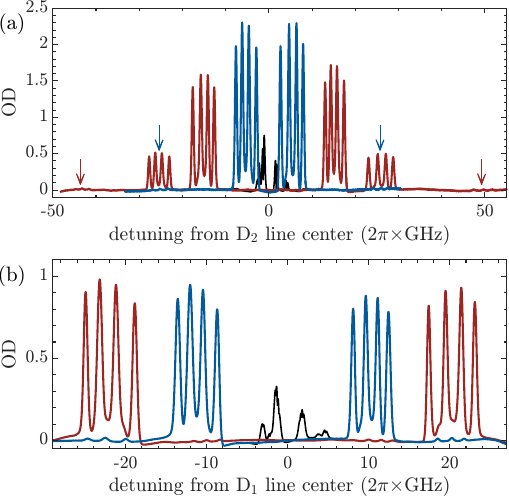}
	\caption{\label{Spectra} Measured spectra of the $^{87}$Rb (a) D$_2$ and (b) D$_1$ lines for an atomic temperature of \SI{92\pm1}{\degreeCelsius}. The blue (red) traces show the spectrum for horizontally (vertically) linearly polarized light in the laboratory frame corresponding to $\pi$ ($\sigma^+$ and $\sigma^-$) transitions. Even though the atomic ensemble is Doppler-broadened, in the HPB regime the single transitions can be individually resolved. At this high temperature the strongest singly forbidden transitions, marked with arrows, can be recognized in the spectra of both lines. The arrows' colors correspond to the trace of interest. In order to give a common sense of scale, the spectrum of a reference cell filled with natural Rb, outside of the magnetic field, is added as black trace (amplitude arbitrarily scaled). Note the difference in horizontal ranges.}
\end{figure}

\begin{figure}
	\includegraphics[scale=1]{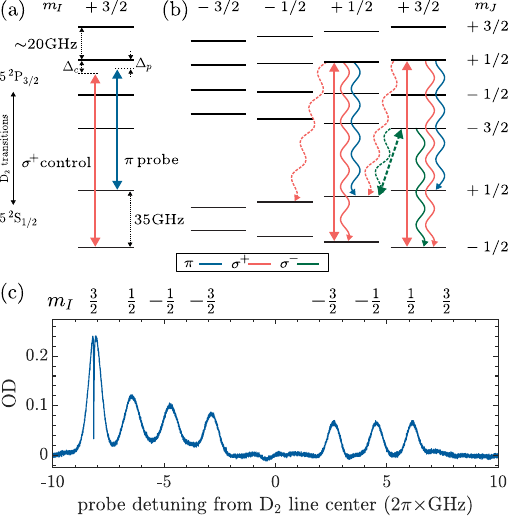}
	\caption{\label{Lambda} (a) Energy scheme of the total electron angular momentum levels of the $m_I = \frac{3}{2}$-manifold. The $\Lambda$ scheme used for the investigation of EIT and ATS in the HPB regime is shown. The detunings of the probe and the control field are labeled $\Delta_p$ and $\Delta_c$, respectively. (b) Level scheme showing the scheme applied to polarize the nuclear spin of the ensemble.  The transitions driven by the pump lasers are drawn as straight lines, differentiating between allowed (solid) and singly forbidden (dashed) transitions. The radiative decay channels are depicted by undulated lines. (c) Spectrum of the eight allowed $\pi$-polarized transitions while the control beam is turned on with a cw power of \SI{4.5}{\milli\watt}. EIT can be observed in the first peak, while the eighth peak  from the absorption spectra above is not present since the control depletes the ground state it addresses. The lower (higher) frequency manifold corresponds to the transitions involving the $m_J = \frac{1}{2}$ ($-\frac{1}{2}$) ground states, while the labels on top indicate the $m_I$ values of the ground states involved in the respective transitions. These data correspond to the lowest control power setting in Fig.~\ref{EITtoAT}(a).} 
\end{figure}

A simple spectroscopic measurement of the Rb vapor in an external field of \SI{1.06}{\tesla} already reveals that single transitions can be resolved in the Doppler broadened medium. The large energy splitting as well as the decoupling of $\mathbf{I}$ and $\mathbf{J}$ allow us to distinguish the hyperfine structure, even of the D$_2$ excited state, without needing to cancel the Doppler effect as in saturated absorption spectroscopy.   
The recorded Rb D line spectra are shown in Fig.~\ref{Spectra}.
Panel (a) depicts the spectrum of the $^{87}$Rb D$_2$ line. The blue trace shows the $\pi$-transitions, which are probed with a horizontally polarized laser. The $\sigma^+$ and $\sigma^-$ transitions, represented in red, are all driven by vertically polarized light. The asymmetries and incomplete transmission between the lines constituting one manifold are mostly due to residual $^{85}$Rb from imperfect isotopic enrichment. 
In all cases the strongest singly forbidden transitions can be recognized.
The full spectrum is manually stitched together from three separate measurements per polarization. In fact, the shown frequency range is larger than the tuning range achieved by a current scan of the DFB laser, and its operating temperature had to be changed to cover the whole range. Horizontally and vertically polarized probes are recorded separately. A Doppler-free saturation spectroscopy in a reference cell outside of the magnetic field serves as absolute frequency reference. This zero field spectrum covers a relatively small frequency range compared to the HPB spectra, making the frequency calibration most reliable near zero detuning. 
In order to account for frequency-dependent power variation, the transmission through the unheated vapor cell is recorded as well. 

In Fig.~\ref{Spectra}(b) the spectrum of the $^{87}$Rb D$_1$ line is shown for the same conditions. Here a commercial \SI{795}{\nano\meter} DFB laser serves as probe. It too is brought to the optical setup through the probe's fiber coupler [label $P$ in Fig.~\ref{Setup}(d)].
Due to the reduced number of possible $m_J$-values, only four allowed transitions each for $\sigma^+$ and $\sigma^-$ are present in the spectrum.
Also note that the frequency splitting between the two $\pi$-polarized quadruplets is larger compared to the D$_2$ line. This fact is due to the smaller Landé $g$-factor of the $5\,^2\text{P}_{1/2}$ term, which results in a smaller energy splitting of the excited states.
As the D$_1$ line covers a narrower frequency range than the D$_2$ line, it is possible to record the entire spectrum with a single scan per polarization. 

\section{\label{sec:EITtoAT}From EIT to ATS}
\begin{figure}
	\includegraphics[width=\columnwidth]{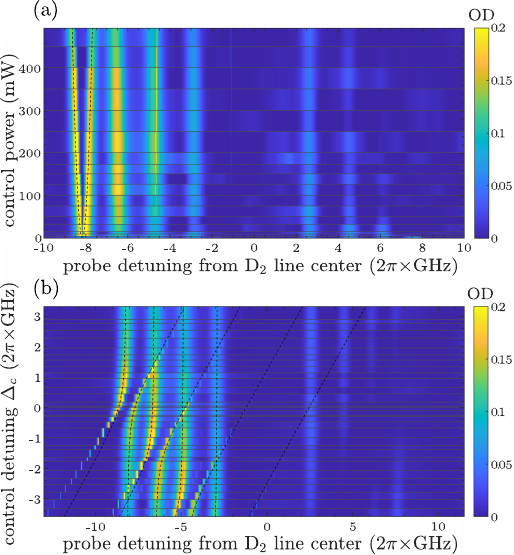}
	\caption{\label{EITtoAT} (a) Measured probe absorption as a function of probe detuning for different control powers showing the transition from EIT to the AT regime. The maximum splitting has a width of approximately \SI{1}{\giga\hertz}. The dashed lines represent a square root fit of the induced transparency width. (b) Measurement of the probe absorption as a function of the probe detuning for different control detunings. The structure of three avoided crossings, the typical signature for strong coupling, can be recognized, while the tail of a fourth anti-crossing can be seen. The dashed lines represent the hyperbolas expected for a three level system. The control detuning is defined with respect to the $\sigma^+$ transition coupling to the $\ket{m_J = -\frac{1}{2},m_I = \frac{3}{2}}$ ground state. In both plots the horizontal lines represent the sampled control powers and detunings, respectively.}
\end{figure}

We studied EIT in order to assess how well a three-level system can be isolated in the HPB regime within the more complex Rb D$_2$ level scheme. Transparency is induced in the ensemble by adding the control beam.
The $\Lambda$ scheme we used for this purpose is illustrated in Fig.~\ref{Lambda}(a).
We chose the probe to be $\pi$-polarized and near resonance with the $\ket{m_J = \frac{1}{2},m_I = \frac{3}{2}} \leftrightarrow \ket{m'_J = \frac{1}{2},m'_I = \frac{3}{2}}$ transition. The control ``leg'' of the $\Lambda$ scheme is consequently chosen to be the $\sigma^+$ transition $\ket{m_J = -\frac{1}{2},m_I = \frac{3}{2}} \leftrightarrow \ket{m'_J = \frac{1}{2},m'_I = \frac{3}{2}}$.
Figure~\ref{Lambda}(c) shows the spectrum resulting from a scan of the probe frequency around the eight allowed $\pi$-polarized transitions while the control frequency is kept fixed. The measurement was performed at an atomic temperature of roughly \SI{47}{\degreeCelsius} in order to maintain sufficient dynamic contrast where the absorption occurs and not to distort the transition lines. 
A deep EIT feature is generated in the first absorption peak. Notably, high transparency is induced in the atomic vapor, almost reaching the transmission level through the ensemble far from resonance, and indeed it becomes unambiguously complete at higher control powers. Even in cold atoms complete transparency is only achievable in certain level-schemes \cite{Yan2001}, in particular only in $\Lambda$-schemes. In hot vapor a ``clean'' three-level system where the degree of control is high enough to optically address individual transitions is required to avoid masking EIT windows \cite{Fulton1995}.    
Furthermore, note that the ground state addressed by the control field is efficiently depleted by optical pumping, resulting in a ``missing'' $\pi$ transition and in an increase in the absorption for the transition from the populated ground state. 

The width of the induced transparency window, as well as the underlying physical process, varies as a function of the applied control power. 
At low control powers, destructive interference between absorption pathways induces the transparency.
In this EIT regime, the width of the transparency window is proportional to the square of the control Rabi frequency, i.e., $\delta\omega_{\text{EIT}}\propto\lvert\Omega_c\rvert^2$ \cite{Fleischhauer2005}.

For high control powers, on the other hand, the transparency is caused by dressing of the states in the strong coupling regime. Transparency manifests as a gap between two separated absorption lines. In this ATS scenario, the width of the transparency is directly proportional to the control Rabi frequency $\Omega_c$ \cite{CohenTannoudji1996}.

Figure~\ref{EITtoAT}(a) shows the absorption profile of the atoms as a function of the probe detuning from the D$_2$ line center for different control powers. The absorption profile transitions smoothly from the EIT regime to ATS, as expected for a textbook three-level system.
Indeed, an initially narrow transparency window evolves into two distinct and well separated peaks with increasing control power, and at the maximal power of \SI{496\pm15}{\milli\watt}, a splitting of approximately \SI{1}{\giga\hertz} is reached.
The consistent absence of the eighth spectral line illustrates that the control field depletes the ground state it addresses even at low powers [cf. Fig.~\ref{Lambda}(c)].
Furthermore, with increasing power, the control field starts pumping the adjacent ground-state levels of the $m_J=-\frac{1}{2}$ manifold as well.
Well into the AT regime, the splitting, defined as the difference between the two absorption maxima, equals the Rabi frequency $\Omega_c$.
As $\Omega_c\propto \sqrt{P}$, the dashed lines in the figure fit the measured splitting with a square root function.

By varying the control frequency we can investigate distinct three-level systems and individually address them. 
We scan the detuning of the control field with respect to the $\sigma^+$ transition coupling to the $\ket{m_J = -\frac{1}{2},m_I = \frac{3}{2}}$ ground state over several gigahertz.
Experimental parameters are kept as in the previous measurement, with \SI{485}{\milli\watt} (near maximum) control power.
The data presented in Fig.~\ref{EITtoAT}(b) show avoided crossings, the typical signature of dressed states \cite{CohenTannoudji1996}. 

For zero control detuning the minimal splitting is achieved and the scenario from Fig.~\ref{EITtoAT}(a) is reproduced, while at larger detunings the splitting increases.
In the asymptotic limit, the doublet is composed of a large peak at the bare transition frequency and a smaller peak approximately moving along a line of unity slope in the control frequency. The amplitude of this second peak decreases with detuning. Within the plotted control detuning range, the control field becomes resonant with three out of four $\sigma^+$ transitions coupling to $m_J = - \frac{1}{2}$ levels, closing $\Lambda$-systems with the probe as it is scanned.
Thus, three avoided crossings and the asymptotic tail of a fourth one can be identified in Fig.~\ref{EITtoAT}(b). If the control frequency were red detuned further, the same behavior would be expected for the probe transitions coupling to the $m_J = -\frac{1}{2}$-manifold.

Each avoided crossing is modeled as an ideal three-level system by the two branches of a hyperbola given by $\Delta_{\pm} = \frac{1}{2}\left( \Delta_c \pm \sqrt{\Delta_c^2 + \Omega_c^2} \right)$, represented as dashed lines in Fig.~\ref{EITtoAT}(b). 
This simple model is in good agreement with the data, illustrating that the various $\Lambda$-schemes addressed by the lasers can be treated as separate three-level systems.
When both the control and probe fields are tuned to the bare transition frequencies, the minimum splitting, corresponding to $\Omega_c$, is achieved. In the model we set $\Omega_c = 2\pi\times\SI{950}{\mega\hertz}$. Along the horizontal axis, which is calibrated with Doppler-free saturation spectroscopy, the hyperbolas are centered around the corresponding, unperturbed probe frequency. The vertical offset is chosen by maximizing the overlap of the data and the model. 
The model diverges from the data at low frequencies, most likely due to a nonlinear frequency scan of the laser. In fact, all available absolute frequency calibration points are close to the central region of the probe frequency scan, deteriorating the reliability of the horizontal axis at the edges of the scan range. 

\section{\label{sec:Ipump}Nuclear Spin Pumping}

In the HPB regime the decoupling of $\mathbf{J}$ and $\mathbf{I}$ divides the atoms into $2I+1$ separate manifolds, which are well isolated from each other. 
Furthermore, the induced energy shifts, mainly of the ground states, result in distinct transition frequencies for the equivalent optical transitions within different manifolds. The frequency of an incoming radiation field can thus be tuned to be resonant with and therefore address atoms within a single nuclear spin manifold.

Although a magnetic field above \SI{1}{\tesla} puts the atomic ensemble well into the HPB regime by Eq.~\eqref{eq:HPBregime}, some coupling of the ground state sublevels persists at such fields, with superposition probability amplitudes on the order of a few percent (see Ref.~\cite{Mottola2023}). Under our conditions it is, therefore, possible to exploit singly forbidden transitions to partially polarize the nuclear spin of the atoms.
In the dark, all ground states in atomic vapors are equally populated due to their thermal population distribution.
In addition to standard optical pumping on dipole-allowed transitions within a nuclear spin manifold, atoms can be transferred between them by driving these forbidden transitions. Thus produced population imbalances increase the optical depth of a selected manifold for a fixed atomic temperature. This is advantageous for processes like coherent absorption, which rely on optical depth for efficiency but are disturbed by collisions \cite{Manz2007}. With increasing field strength, the viability of this approach diminishes; however, fuller decoupling also ensures that, for whatever interaction is under study, atoms in ``wrong'' $m_I$ manifolds effectively act as a mere background gas, which naturally mitigates the need for highly polarized ensembles with regards to the suppression of spurious optical processes.

DFB diodes, available at the Rb wavelength, and digital controllers for butterfly laser diodes constitute affordable and ready-to-use/plug-and-play devices, which allow for an easy state preparation with dedicated pump lasers for each different pump transition.
Using dedicated pump lasers avoids involuntarily creating coherences between the optical fields, which could inadvertently prepare the atoms in dark states \cite{Pandey2013}.

For a proof-of-principle test of a combined pumping scheme, as shown in Fig.~\ref{Lambda}(b), three pump lasers were utilized. The primary pump drives the same transition as the control in the previous sections (depleting the $\ket{m_J = -\frac{1}{2},m_I = \frac{3}{2}}$ state with $\sigma^+$ polarized light). A further DFB laser addresses the singly forbidden $\sigma^-$ transition coupling the states $\ket{m_J = \frac{1}{2},m_I = \frac{1}{2}} \leftrightarrow \ket{m'_J = -\frac{3}{2},m'_I = \frac{3}{2}}$ (corresponding to the leftmost transition in Fig.~\ref{TransitionsD2}). This forbidden pump transfers atomic population from $m_I = \frac{1}{2}$ to $m_I = \frac{3}{2}$.
Finally, a third pump laser drives the allowed $\sigma^+$ transition coupling the states $\ket{m_J = -\frac{1}{2},m_I = \frac{1}{2}} \leftrightarrow \ket{m'_J = \frac{1}{2},m'_I = \frac{1}{2}}$, which repopulates the state addressed by the forbidden pump.
The two additional pump lasers are aligned under a small angle with respect to the counter-propagating primary pump by using D-shaped mirrors (not shown in the setup sketch). All three pump lasers were operated at an optical power of \SI{20}{\milli\watt}, limited by the damage thresholds of the fiber-based, fast optical switches intended for future experiments.

\begin{figure}
	\includegraphics[scale=1]{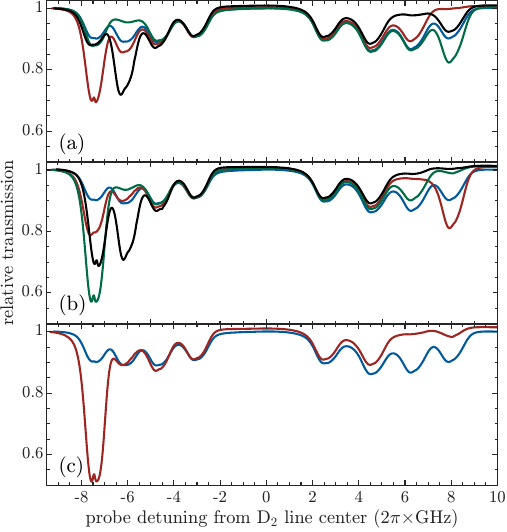}
	\caption{\label{pumping} Experimental implementation of the (partial) pumping scheme involving allowed as well as ``singly forbidden'' transitions. The measurements were performed at an approximate applied magnetic field of \SI{1.06}{\tesla}. In all three panels the spectrum of the unpumped atoms is shown in blue as a reference. The background of the reference spectrum is modeled as a second-order polynomial used to correct all shown traces. (a) The effect of each individual pump laser on the atoms is shown: $\sigma^+$ polarized  $m_I=\frac{3}{2}$-pump (red), or $\sigma^-$ polarized forbidden pump (green), or $\sigma^+$ polarized $m_I=\frac{1}{2}$-pump (black). (b) The effect of two lasers combined is shown. The effects of the forbidden pump and the $m_I=\frac{1}{2}$-pump are depicted in red. The green trace shows the spectrum obtained by using the $m_I=\frac{3}{2}$-pump and the forbidden pump. Turning on both allowed pumps ($m_I=\frac{1}{2}$ and $m_I=\frac{3}{2}$) results in the black trace. (c) All three pump lasers are used (red), as depicted in the level-scheme in Fig.\ref{Lambda}(b).}
\end{figure}

Figure~\ref{pumping} shows the spectrum of the probe, which is scanned over all allowed $\pi$ transitions, under different pumping conditions.
An unpumped spectrum (blue) of the probe is added to each panel for comparison.
In order to correct for frequency dependent changes of the emitted probe power, the background of the unpumped spectrum is fitted with a second-order polynomial and used to correct all traces shown in the figure.

In Fig.~\ref{pumping}(a) the spectra for the scenarios in which only one pump laser is on are shown. The ground state addressed by the corresponding pumper is depleted, resulting in uniformly high transmission over the full line width of the respective transitions. The combined effect of each possible pair of pumping lasers applied together is plotted in Fig.~\ref{pumping}(b); see caption for details. Finally, Fig.~\ref{pumping}(c) shows the spectrum obtained when all three pump lasers are on. In this scenario, as expected, we observe the strongest absorption for the $\ket{m_J=\frac{1}{2},m_I=\frac{3}{2}}\leftrightarrow\ket{m_J'=\frac{1}{2},m_I'=\frac{3}{2}}$ $\pi$-transition, indicating that the yielded optical depth is higher than what is possible by solely pumping within $m_I=\frac{3}{2}$ [cf. red trace in Fig.~\ref{pumping}(a)]. 

It can be noted that in Fig.~\ref{pumping}(c) the absorption of the $\ket{m_J = \frac{1}{2},m_I = \frac{1}{2}}$ state, addressed by the forbidden pump,  appears to be unchanged compared to the unpumped spectrum. This is already the case in the data presented in red in Fig.~\ref{pumping}(b), where the pump lasers only address atoms with $m_I = \frac{1}{2}$. This indicates the presence of non-radiative nuclear spin relaxation processes, e.g., wall collisions, which compete with nuclear spin pumping.

From the measured spectra it appears that the $\ket{m_J=-\frac{1}{2},m_I=\frac{1}{2}}$ ground state is depleted while $\ket{m_J=\frac{1}{2},m_I=\frac{1}{2}}$ shows no significant difference with respect to the reference curve. We thus estimate that the implemented pumping scheme transfers about \SI{50}{\percent} of the atomic population of $m_I=\frac{1}{2}$ to  $m_I=\frac{3}{2}$. The presented pumping scheme can be expanded by pumping each manifold into the respective $m_J = \frac{1}{2}$ state and by using the forbidden transitions to transfer the atomic population towards manifolds of higher $m_I$-values. With such an expansion, under the presented experimental conditions, and assuming that each manifold can be half depleted, it should thus be possible to achieve a total atomic polarization of $\frac{15}{32}$, despite the thorough isolation of the nuclear states from one another.

\section{\label{sec:SummaryAndOutlook}Summary and Outlook}
In this article we have investigated EIT and optical pumping in warm Rb vapor in the HPB regime. We have demonstrated the isolation of a ``clean'' three-level $\Lambda$-system in a strong external magnetic field, and we have explored the atomic polarizability both within and across nuclear spin manifolds under these operating conditions. Furthermore, we have studied the phenomena of EIT and AT line-splitting in this scheme, reproducing spectra matching the textbook ideal in typically far more complicated hot alkali-metal vapor. Both aspects are of particular interest for the implementation of a broad class of quantum technologies in such hot vapors. These include, for instance, optical ground-state quantum memories, as memory efficiency is a function of optical depth \cite{Gorshkov2007} and an ideal three-level system suppresses noise processes \cite{Manz2007,Lauk2013} and undesirable interference \cite{Wolters2017} that take place in the presence of more energy levels. Further, bi-photon generation schemes in ensembles, such as those based on spontaneous four wave mixing, would not only be realizable at the investigated optical depths (compare, e.g., Ref.~\cite{Davidson2021} and references therein) but would also cover an atypically large range of frequencies for atomic sources by transition selection and magnetic field tuning.

In summary our investigation yields positive prospects for putting hot atomic ensembles deep in the HPB regime to work in applications that require both well isolated $\Lambda$ systems and moderate to high optical depths. Indeed, in a concurrent article \cite{Mottola2023PRL}, we perform a proof-of-principle quantum memory experiment in this system, demonstrating low noise for storage at the single-photon level and an internal efficiency at the theoretical limit.

\begin{acknowledgments}
The authors thank Gaetano Mileti for supplying us with the microfabricated vapor cell as well as Florian Gruet who helped us characterize the cell's content. We acknowledge financial support from the Swiss National Science Foundation through NCCR QSIT and from the European Union through the Quantum Flagship project macQsimal (Grant No.
820393).
\end{acknowledgments}

\end{document}